\documentclass[useAMS,usenatbib]{mn2e}

\usepackage{graphicx}
\usepackage{breqn}

\title[Gap opening beyond dead zones]{Gap opening beyond dead zones by photoevaporation}
\author[R. Morishima]{R. Morishima$^{1,2}$\thanks{E-mail:
ryuji.morishima@jpl.nasa.gov}\\
$^{1}$University of California, Los Angels, California, USA\\
$^{2}$Jet Propulsion Laboratory, Pasadena, California, USA}
\begin{document}

\date{Accepted 2011 September 30. Received 2011 June 28.}

\pagerange{\pageref{firstpage}--\pageref{lastpage}} \pubyear{2011}

\maketitle

\label{firstpage}

\begin{abstract}
We propose a new hypothesis for the origin of protoplanetary discs with large inner holes (or gaps), so-called transition discs. 
Our gas disc model takes into account layered accretion, in which poorly-ionized low-viscosity dead zones 
are sandwiched by high-viscosity surface layers, 
and photoevaporative winds induced by X-rays from the central stars. We find that a gap opens at a radius outside a dead zone,
if the mass loss rate due to photoevaporative winds exceeds the mass accretion rate in the dead zone region. 
Since the dead zone survives even after the gap opens, mass accretion onto the central star continues for a long time.
This model can reproduce large gap sizes  and high mass accretion rates seen in observed transition discs.

\end{abstract}


\section{Introduction}

Spatial distribution of dust in protoplanetary discs is measured from disc continuum spectra in
near-infrared to millimeter wavelengths
 whereas the gas density is measured directly by line emission from such as CO and indirectly from H$\alpha$ emission which 
is used for estimations of accretion rates of gas onto central stars.  
Infrared excess due to dust emission disappears in 3 Myr for roughly a half of discs and for almost all discs in 10 Myr \citep{Hai01}.
Gas in inner discs disappears in similar time-scales  \citep{Fed10}, but the life time of gas in outer discs is not clearly known.

Recent observations,  such by the Spitzer infrared space-telescope, the Submillimeter Array, and the Subaru telescope,  revealed discs in the dispersal phase,
so-called transition discs  \citep{Mer10, Tha10, And11}. They have optically thick outer discs whereas inner discs with sizes up to $\sim$ 70 AU 
are optically thin. This indicates depletion of dust  in the inner discs. Some of transition discs show near-infrared excess indicating formation of 
gaps instead of holes and some others have weaker mid to far infrared emission indicating optically thinner outer discs  \citep{Muz10}.
The fraction of transition discs in protoplanetary discs increases with time \citep{Cur11}: 15-20 per cent at 1-2 Myr and more than 50 per cent at 5-8 Myr.
A puzzling thing is that a large fraction (75 per cent; \citet{Mer10}) 
of transition discs exhibit gas accretion onto their central stars and their accretion rates are close to those for classical T Tauri stars, 
 $\sim 10^{-8}$ $M_{\odot}$ yr$^{-1}$ \citep{Har98}. 

There are three proposed explanations for the origin of transition discs but all of them do not seem to be fully satisfactory \citep{And11,Wil11}. 
(1) Gap opening by photoevaporative winds (Gorti, Dullemond \& Hollenbach 2009; Owen, Ercolano \& Clarke 2011).   
If the mass loss rate of photoevaporative winds
exceeds the mass accretion rate of viscous evolution, a gap can open. However, before a gap opens, the mass accretion rate 
largely decreases and becomes much less than $10^{-8}$ $M_{\odot}$ yr$^{-1}$. This is the case even with X-ray 
photoevaporation with the mass loss rate as high as $10^{-8}$ $M_{\odot}$ yr$^{-1}$ \citep{Owe11}.
 The sizes of the inner holes due to photoevaporation
cannot be larger than 20 AU when mass accretion onto the central star still remains. 
(2) Dust growth (Dullemond \& Dominik 2005;  Tanaka, Himeno \& Ida 2005). This is expected particularly in inner discs as the models 
show that the growth time of dust has strong dependence on distance from the central star. However, dust growth alone cannot 
be responsible for transition discs, because the models for dust growth predict smooth opacity changes with radius  whereas observed transition
discs show clear gaps of dust. (3) Gap opening by giant planets. \citet{Zhu11}  show that multiple giant planets can open 
a gap as wide as the largest holes observed. However,  depletion of the surface density in the gap results in a large decrease of the mass accretion rate. 

The most probable source of the disc viscosity is the turbulent viscosity caused by the magneto-rotational  instability (MRI;  \citet{Bal91}).
Since the ionization rate in the thick inner disc is not high enough, a poorly ionized layer with a low viscosity called a dead zone forms near the disc midplane 
sandwiched by MRI active layers (Gammie 1996; Zhu, Hartmann \& Gammie 2010). This type of accretion is named layered accretion.
In this paper,  we develop a gas disc model which takes into account layered accretion and photoevaporative winds. 
Layered accretion has not been taken into account in previous works of photoevaporation.
As we will show, our disc model can reproduce some of important properties of observed transition discs,
such as large gap sizes and high mass accretion rates. 
 
 In Sec.~2, we introduce our gas disc model.
 In Sec.~3, results are shown for both discs with and without layered accretion over wide ranges of parameters.
 In Sec.~4, implications for transitions discs are discussed.
 In Sec.~5, the summary of this study is given.

\section{Methods}
\subsection{Basic equations}
We assume that mass accretion from 
a molecular cloud core has already completed and that 
 the mass of the central star is fixed during disc evolution.
Time evolution of the surface density $\Sigma$ is derived from 
the mass and angular momentum conservation equations:
\begin{equation}
\frac{\partial \Sigma}{\partial t} = \frac{1}{2 \pi r} \frac{\partial \dot{M}}{\partial r} -  \dot{\Sigma}_{\rm w}, 
\label{eq:massc}
\end{equation} 
\begin{equation}
\frac{\partial (\Sigma j)}{\partial t} = \frac{1}{2 \pi r} \frac{\partial}{\partial r}\left(\dot{M} j + \dot{J}_{\rm vis} \right) -  \dot{\Sigma}_{\rm w}j, 
\label{eq:angc}
\end{equation}
where $\dot{M}$ is the mass accretion rate (assumed to be positive for inward accretion), 
$\dot{\Sigma}_{\rm w}$ is the mass loss flux of photoevaporative winds,  
$j = r^2\Omega$ is the specific angular momentum with $\Omega$ being the Keplerian frequency, 
and $\dot{J}_{\rm vis} = 2 \pi r^3 \Sigma \nu d\Omega/dr$ is the angular momentum flux transferred by the disc viscosity, $\nu$. 
From Eqs.~(\ref{eq:massc}) and (\ref{eq:angc}), the mass accretion rate is obtained as  
\begin{equation}
\dot{M}= -2 \pi r \Sigma v_{r} = 6 \pi r^{1/2} \frac{\partial}{\partial r} (\Sigma\nu r^{1/2}), \label{eq:mt}
\end{equation}
where $v_{r}$ is the radial velocity. Substituting this form to Eq.~(\ref{eq:massc}), we obtain 
the well known diffusion equation \citep{Lyn74}: 
\begin{equation}
\frac{\partial \Sigma}{\partial t} = \frac{3}{r}\frac{\partial}{\partial r}\left[ r^{1/2} \frac{\partial}{\partial r} (\Sigma\nu r^{1/2}) \right] -\dot{\Sigma}_{\rm w}.
\label{eq:sigd}
\end{equation}

\subsection{Disc viscosity}
Using the viscosity parameter $\alpha$ \citep{Sha73}, the viscosity is given as 
\begin{equation}
\nu = \alpha c h  = \frac{\alpha R T}{\mu \Omega}, \label{eq:nu}
\end{equation}
where $c $ is the isothermal sound velocity at the disc midplane 
and $h=c/\Omega_{\rm K}$ is the  vertical scale height of the disc. The isothermal sound velocity is 
given as  $c = (RT/\mu)^{1/2}$, where $R$ is the gas constant,
$T$ is the disc temperature at the midplane, and we take the mean molecular weight as $\mu$ = 2.33 g mol$^{-1}$.
For models without dead zones,  $\alpha$ is radially constant. For models with dead zones,  the $\alpha$ parameters for active surface layers
and dead zones are defined to be $\alpha_{\rm a}$ and  $\alpha_{\rm d}$. 
Then, the effective disc viscosity parameter is given as 
\begin{equation}
\alpha = \frac{ \Sigma_{\rm a} \alpha_{\rm a} + (\Sigma-\Sigma_{\rm a})\alpha_{\rm d}}{\Sigma}, \label{eq:ald}
\end{equation}
where $\Sigma_{\rm a}$ is the sum of the surface densities of both sides of active layers. 
In simulations, we adopt $\alpha_{\rm a} = 0.01$.
The surface density $\Sigma_{\rm a}$ is modeled after Suzuki, Muto \&  Inutsuka (2010):
\begin{equation}
\Sigma_{\rm a} =   {\rm min}\left[\Sigma_{\rm CR}+\Sigma_{\rm X}\left(\frac{r}{\rm 1 AU}\right)^{-2}, \Sigma \right], \label{eq:siga}
\end{equation}
where $\Sigma_{\rm CR}$ is the surface density of layers ionized by cosmic rays and  $\Sigma_{\rm X}$ is the surface density 
of layers ionized by X-rays from the central star at 1 AU. 
Equations~(\ref{eq:ald}) and (\ref{eq:siga}) indicate that $\alpha \ll \alpha_{\rm a}$ if $\Sigma \gg \Sigma_{\rm a}$.
On the other hand, if $\Sigma \le \Sigma_{\rm CR}+\Sigma_{\rm X}(r/{\rm 1 AU})^{-2}$, 
the region is entirely MRI active ($\Sigma = \Sigma_{\rm a}$) so that $\alpha = \alpha_{\rm a}$.

The surface densities  $\Sigma_{\rm CR}$ and $\Sigma_{\rm X}$ 
particularly depend on dust amount in the disc, as recombination of ions and electrons
occurs on the dust surface \citep{San00,Bai11}.
For example,  \citet{Suz10} uses $\Sigma_{\rm CR}$ = 12 g cm$^{-2}$ and $\Sigma_{\rm X}$ = 25 g cm$^{-2}$,
whereas \citet{Gam96} adopts nominal values as   $\Sigma_{\rm CR}$ = 200 g cm$^{-2}$ and $\Sigma_{\rm X}$ = 0.
The value of $\Sigma_{\rm CR}$ adopted in  \citet{Gam96} comes from the attenuation length of comic rays ($\simeq$ 96 g cm$^{-2}$; \citet{Ume81}).
If the dust abundance is negligible, MRI is sustained in the layers as thick as the attenuation length \citep{San00}.
On the other hand, in the MHD simulations of \citet{Suz10}, they adopt the ionization degree calculated in 
 \citet{San00} and \citet{Inu05} assuming that the dust-to-gas ratio be 0.01.
In the present study, we vary $\Sigma_{\rm CR}$  as a parameter. On the other hand,
we fix $\Sigma_{\rm X}$ to be 25 g cm$^{-2}$, as we find that $\Sigma_{\rm X}$ is less important than 
$\Sigma_{\rm CR}$ for overall disk evolution, in particular, for $\alpha_{\rm d} \ga 10^{-5} $.
MHD simulations show that dead zones have non-zero viscosities \citep{Fle03,Suz10}; the value of $\alpha_{\rm d}$ 
is on the order of $\sim 10^{-5}$-$10^{-3}$.

\subsection{Disc temperature}

\begin{table}
\caption{Rosseland mean opacity from \citet{Ste98}.}

\begin{tabular}{@{}ccc} \hline
Applicability & $\kappa_0$ (cm$^2$ g$^{-1}$)  & $\beta$\\  \hline
  $T < 150$ K & $2 \times 10^{-4}$  & 2\\
150 K $< T \le$ 180 K & $1.15 \times 10^{18} $ & -8\\
 180 K $< T \le$ 1380 K & $2.13 \times 10^{-2} $ & 3/4\\
1380 K $\le T$  & $4.38 \times 10^{44} $  & -14\\ \hline
\end{tabular}

\end{table}

As heat sources of discs,  we take into account viscous heating, irradiation from the central star, 
and background irradiation. We simply superpose all contributions which are independently modeled:
 \begin{equation}
T^4 = T_{\rm vis}^4 + T_{\rm irr}^4 + T_{\rm amb}^4,
\end{equation}
where $T_{\rm vis}$, $T_{\rm irr}$, and $T_{\rm amb}$ are the temperatures contributed from viscous heating, 
irradiation from the central star, and background irradiation.
If  only viscous heating is considered, 
the radiative cooling from both sides of the disc balances with the viscous heating as 
\begin{equation}
2 \sigma_{\rm SB} T_{\rm vis,eff}^4 = \frac{9}{4}\Sigma \nu \Omega^2, \label{eq:tveff}
\end{equation}
where $T_{\rm vis, eff}$ is the effective disc surface temperature.
The relation between the midplane temperature $T_{\rm vis}$ and the effective temperature 
is given by \citet{Hub90} as 
\begin{equation}
T_{\rm vis}^4 = \tau_{\rm eff} T_{\rm vis, eff}^4, \label{eq:tvis}
\end{equation}
with
\begin{equation}
\tau_{\rm eff} = 
\left[\frac{3\tau}{8}\left(1+\frac{\Sigma_{\rm a}}{\Sigma}-\frac{\alpha_{\rm a}\Sigma_{\rm a}}{\alpha\Sigma} \right) + 
\frac{\sqrt{3}}{4} + \frac{1}{3\tau}\right], \label{eq:taueff}
\end{equation}
where $\tau = \kappa \Sigma/2$ is the optical depth at the midplane. 
Eq.~(\ref{eq:taueff}) is applicable to all cases. 
For pure MRI active regions without dead zones, we set $\Sigma = \Sigma_{\rm a}$ and $\alpha = \alpha_{\rm a}$.
For the dead zone region with $\alpha_{\rm d} = 0$, we set $\Sigma\alpha  = \Sigma_{\rm a}\alpha_{\rm a}$
so that Eq.~(\ref{eq:tvis}) becomes the same with Eq.~(7) of \citet{Gam96} for $\tau \Sigma_{\rm a}/\Sigma \gg 1$.
A detailed derivation of the first term in the square bracket of the right hand side of Eq.~(\ref{eq:taueff}) is given in \citet{Wun06}.
The opacity $\kappa$ is represented by $\kappa = \kappa_0 T^{\beta}$, where $\kappa_0$ is a constant.
We use $\kappa$ from \citet{Ste98} shown in Table~1.  Results of the present study do not sensitively depend on opacity law.

The disc effective temperature irradiated by the central star is given by \citep{Rud91} 
\begin{equation}
T_{\rm irr, eff}^4 =  T_{*}^4 \left[\frac{2}{3\pi}\left(\frac{r_{*}}{r}\right)^3  + 
\frac{1}{2}\left(\frac{r_{*}}{r}\right)^2\left(\frac{h}{r}\right)\left(\frac{d\ln{h}}{d\ln{r}} -1\right)  \right]\label{eq:tc},
\end{equation}
where $T_{*}$ and $r_{*}$ are the photospheric temperature and radius of the central star. We adopt $T_{*} = 4000$ K and $r_{*} = 3$ $r_{\odot}$.
The scale height $h$ used in Eq.~(\ref{eq:tc}) is the photospheric disc scale height and we simply assume it to be the same with the pressure scale height. 
After \citet{Hue05},   we assume $d\ln{h}/d\ln{r} = 9/7$. This is validated as 
$T \propto r^{-3/7}$ at large radii where irradiation from the central star is usually the dominant heat source.
The midplane temperature,  $T_{\rm irr}$, irradiated by the central star is calculated using the vertical temperature structure derived by \citet{Mal91}
(see also Malbet,  Lachaume \& Monin (2001)):
\begin{dmath}
 T_{\rm irr}^4 = T_{\rm irr, eff}^4\left[\frac{3}{4}\mu_0\left(1-\exp{\left(-\frac{\tau}{\mu_0}\right)}\right) + \frac{1}{2} 
 + \frac{1}{4\mu_0}\exp{\left(-\frac{\tau}{\mu_0}\right)}\right].\label{eq:tirr}
\end{dmath}
Here $\mu_0$ is the averaged cosine of the incident angle of the central star given as 
\begin{equation}
\mu_0 = -\frac{H^0}{J^0}, 
\end{equation}
with the zeroth and first order moments, $J^0$ and $H^0$, of the incident intensity at the upper most layer:
\begin{equation}
J^0 =\frac{\sigma_{\rm SB}T_{*}^4}{4\pi}\left[1-\left(1-\left(\frac{r_{*}}{r}\right)^2\right)^{1/2}\right], 
\end{equation}
\begin{equation}
H^0 = -\frac{\sigma_{\rm SB}T_{\rm irr,eff}^4}{4\pi}.
\end{equation}
Eq.~(\ref{eq:tirr}) represents that $T_{\rm irr}$ of an optically thick disc  
is lower than $T_{\rm irr, eff}$ by  a factor of $2^{1/4}$, as the upper most layers emit 
half of the received energy toward the midplane. In the optically thin limit ($\tau \rightarrow 0$),  the third term in 
the bracket dominates and we obtain 
$T_{\rm irr}^4 = T_{*}^4(r_{*}/r)^2/8$, which corresponds to the equilibrium temperature irradiated by the half hemisphere 
of the central star without any obstacle. The temperature at $\tau \rightarrow 0$ should be higher by a factor of $2^{1/4}$, but 
this difference does not affect our results as the stage with a low surface density is very short with photoevaporative winds.

The ambient temperature is simply assumed as $T_{\rm amb} = 20$ K.

\subsection{Mass loss due to photoevaporation}
Among various types of photoevaporations, the one induced by X-rays from the central star is 
most strong and most important for gas disc evolution in the planet formation region \citep{Erc09, Owe10, Owe11}, 
unless discs are close to very massive external stars \citep{Ada04,Mit10}.
The total mass loss rate from a disc due to X-ray photoevaporation is estimated by \citet{Owe11} as 
\begin{equation}
\dot{M}_{\rm w} = 6.4 \times 10^{-9} A \left(\frac{L_{\rm X}}{10^{30} \hspace{0.3em} {\rm erg s}^{-1}}\right)^{1.14} M_{\odot} \hspace{0.3em}{\rm yr}^{-1},
\end{equation}
where $A$ is a constant of the order of unity (we fix it to be unity) and $L_{\rm X}$ is the X-ray luminosity from the central star. For 
sub-solar to solar mass protostars, $L_{\rm X}$ is $10^{29} $-$10^{31}$ erg s$^{-1}$ \citep{Gud07}.
Numerical simulations of \citet{Owe10} show that $\dot{\Sigma}_{\rm w}$ is 
roughly proportional to $r^{-3/2}$ (their Fig.~13). Thus, we give the mass loss rate per unit area as 
\begin{equation}
\dot{\Sigma}_{\rm w} = \frac{\dot{M}_{\rm w}}{4\pi (r_{\rm X,out}^{1/2}-r_{\rm X,in}^{1/2})} r^{-3/2} \hspace{0.3em} 
({\rm for} \hspace{0.3em} r_{\rm X, in }\le r \le r_{\rm X, out}), 
\end{equation}
where $r_{\rm X, in}$ and $r_{\rm X,out}$ are the inner and outer edges 
of the range where mass loss occurs. 
We take $r_{\rm X,in} =$ 1 AU and $r_{\rm X,out} =$ 70 AU from \citet{Owe10}. 
Outside $r_{\rm X, out}$, other mechanisms such as EUV or FUV photoevaporation \citep{Gor09} 
may work, although these effects are not taken into account in the present work. 
Results of disc evolution do not sensitively depend on $r$-dependence of $\dot{\Sigma}_{\rm w}$,
as similar results are obtained even if we vary  the power-law index between -1 and -2.

In some test simulations, we also adopt the mass loss due to MRI disc winds \citep{Suz10}.
The mass loss rate of MRI disc winds at a given radius is proportional to the surface density.
We find that  gas depletes very rapidly even near  the inner edge, and that the mass accretion rate onto the central star becomes
less than $10^{-9} M_{\odot}$ yr$^{-1}$ within 1 Myr, as shown in \citet{Suz10}. This accretion rate looks too low as compared with those for 
typical discs around classical T Tauri stars \citep{Har98}. Probably, the mass loss rate of MRI disc winds is much smaller, but this mechanism 
may still play an important role in disc evolution. 

 \subsection{Numerical procedures} 
 
  We numerically solve Eq.~(\ref{eq:sigd}) using the method described in Bath and Pringle (1981), in which the radial grid size
is proportional to $\sqrt{r}$. We use 1000 grids between 0.03 AU and 3000 AU. These numbers give the inner most grid size of 0.02 AU. 
At the inner boundary $\Sigma$ is fixed to be zero, and at the outer boundary we adopt the outward mass flux given by  $-3\pi\Sigma\nu$.  
We find that the evolution of the total mass is nearly the same even with larger outer boundary radii whereas artificial mass loss 
from the outer boundary is not negligible with smaller outer boundary radii. The temperature is calculated by the Newton-Raphson method.

For all simulations, the mass of the central star is the solar mass, $M_{\odot}$, 
the initial mass of the disc is 0.1 $M_{\odot}$, and the initial surface density is proportional to $r^{-1}$ with the outer edge of 20 AU. 
The adopted time step size is 1 yr.  A single simulation for $\sim$ 10 Myr takes roughly one cpu day.
We conduct 8 runs for discs with radially constant $\alpha$'s and 18 runs for discs with dead zones. 
The input parameters are shown in Tables~2 and 3 for simulations of discs without and with dead zones.

 \begin{table}
\caption{Input parameters for simulations of discs without dead zones. }
\begin{tabular}{@{}ccc} \hline
Run ID &  $L_{\rm X}$ & $\alpha$   \\  
              &($10^{30}$ erg s$^{-1}$) &   \\ \hline
  N1       & 1.0 &   $10^{-2}$       \\
 N2         &   " & $10^{-3}$          \\
 N3        &   " &$3 \times 10^{-4}$          \\
 N4           & "   &$10^{-4}$     \\
  N5          &  3.0 &$10^{-2}$      \\
 N6          &   " &$10^{-3}$        \\
 N7           &  "  &$3 \times 10^{-4}$         \\
 N8           &  "  &$ 10^{-4}$        \\ \hline
 \end{tabular}

\medskip
\end{table}

\begin{table}

\caption{Input parameters for simulations of discs with dead zones. We adopt $\alpha_{\rm a} = 0.01$.}   
\begin{tabular}{@{}cccc} \hline
Run ID &  $L_{\rm X}$ & $\Sigma_{\rm CR}$ & $\alpha_{\rm d}$   \\  
              &  ($10^{30}$ erg s$^{-1}$)  &  (g cm$^{-2}$)    &       \\       \hline
 D1     &  1.0         &  50  &  0 \\   
 D2     &  "         &  100   &  " \\
 D3     &  3.0         &  50 &  " \\   
 D4     &  "        &  100  &  " \\ 
 D5        &   1.0   & 25   &$10^{-5}$\\
 D6       &    "       &  50   &  "  \\
 D7       &   "        &  100     &  "  \\
 D8       &   3.0    &  25     &  "  \\
 D9       &   "       &  50     &  "  \\
 D10       &   "        &  100    &  "   \\
 D11      &   "        &  200    &  "   \\
  D12       &   0.3       &  25    &  "   \\
 D13      &   "        &  50    &  " \\
 D14       &  "     &  100     &  " \\ 
 D15     &  1.0     &  50      &  $10^{-4}$ \\   
 D16     &  3.0     &  "        &  " \\
 D17     &  1.0      &  "       &  $10^{-3}$  \\   
 D18     &  3.0      &  "      &  "   \\     \hline
 \end{tabular}
 
\medskip

\end{table}

\begin{table*}
\caption{Summary of disc properties.}
\begin{tabular}{@{}ccccccccc} \hline
Run ID & $\dot{M}_{*, {\rm 1 Myr}}$ &  $t_{\rm d}$  & $t_{\rm dz}$ &  $t_{\rm gap}$ & $r_{\rm gap}$ & $M_{\rm gap}$ & $M_{\rm gap,out}$ & $\dot{M}_{*, {\rm gap}}$\\  
              &($10^{-9} M_{\odot}$ yr$^{-1}$) & (Myr)  & (Myr) &  (Myr) &  (AU)   & ($10^{-3} M_{\odot}$)& ($10^{-3} M_{\odot}$) & ($10^{-9} M_{\odot}$ yr$^{-1}$)\\       \hline
 N1       &7.73& 2.40     &-&2.40&1.47 & 10.23 & 10.23& 5.27  $\times 10^{-4}$ \\
 N2       &9.19&6.12     &-&6.01&1.34 & 10.72 & 10.72 & 2.58 $\times 10^{-2}$\\
 N3       &6.73&10.21        &-&9.08&60.37 & 6.85 & 1.25& 0.17  \\
 N4        &3.22 &16.02   & -& 14.90 & 1.34 & 2.65 & 2.61 &3.21 $\times 10^{-2}$ \\
 N5       &1.27&1.11 &-&1.09&1.22 & 12.63 & 12.62 & 5.86 $\times 10^{-2}$\\
 N6       &6.80 &2.84 &-&2.63&60.37 & 5.46 & 1.37 & 0.20 \\
 N7       &5.31 &4.58  &-&4.12 &1.34 & 5.62 & 5.59 & 7.99  $\times 10^{-2}$\\
 N8       &2.56 &6.46   &-&5.12&1.34 & 14.29 &  14.21 & 7.12 $\times 10^{-2}$ \\  \hline
 D1     &2.89    &12.17       &12.11 &4.36&47.55 & 47.37 & 2.87 & 2.89  \\  
 D2      &5.61   & 8.01     &7.92 &3.71&31.18 & 34.79 & 4.77 & 5.51  \\
 D3        &2.89   & 6.58   &6.54&1.48&54.58 & 57.09 & 2.12 & 2.89 \\
 D4        &5.61   & 5.08   &5.03&1.66 &42.42& 44.84 & 3.94 & 5.51  \\
 D5       &3.48 & 9.39    &9.39& 5.52 & 57.02 & 34.43 & 1.85 &4.40 \\
 D6        &6.89 & 6.24   &6.18&5.37 & 45.32  & 10.98 & 3.17 & 5.80  \\
 D7        &18.31 & 3.69    &3.46 &3.68 & 2.05  & 7.11 & 7.11 & 1.74 $\times 10^{-3}$ \\
 D8        &3.26  &5.40     &5.37& 1.30 &61.23 & 63.56 & 1.03 & 3.22 \\
 D9        &6.53   &3.97     &3.94& 1.62 &57.02 & 48.91 & 2.10 & 7.28  \\
 D10       &17.71   &2.51     &2.47& 1.83 &45.32 & 20.08 & 4.10 &15.88  \\
 D11       &30.85  &1.50   &1.37& 1.49 &2.54 & 8.19 & 8.19& 1.89 $\times 10^{-2}$ \\
  D12       &3.55     &14.32    &14.15&14.29 & 28.20 & 1.66 & 1.65 & 4.51 $\times 10^{-2}$ \\
 D13       &7.00    &9.01     &8.28&9.00&1.22 & 4.32 & 4.32 & 2.24  $\times 10^{-3}$ \\
 D14      &18.49   &6.50     &4.07&6.49& 1.10 & 5.75 & 5.75 & 1.69 $\times 10^{-3}$\\ 
 D15        &17.37  &4.10    &3.95&4.09&2.05 & 6.40  & 6.40 & 5.77 $\times 10^{-3}$ \\
 D16        &15.84   & 2.62 &2.58 &2.02& 52.18 & 14.87 &  2.98 &10.00  \\
 D17       &19.99   &2.83    &2.20 &2.83& 1.47 & 9.51 & 9.51 & 8.74  $\times 10^{-4}$ \\
 D18       &15.31    &1.56   &1.48&1.55& 2.72 & 8.17 & 8.17 & 1.23 $\times 10^{-2}$ \\ \hline
  \end{tabular}
  
\medskip

 $\dot{M}_{*, \rm 1 Myr}$ is the mass accretion rate onto the central star at $t=1$ Myr, $t_{\rm d}$ is the time when $\dot{M}_{*}$
  becomes  $10^{-12}$ $M_{\odot}$ yr$^{-1}$, $t_{\rm dz}$ is the life time of the dead zone (when a disc becomes entirely MRI active for all radii), 
  $t_{\rm gap}$ is the time when a gap opens, $r_{\rm gap}$ is the gap radius where $\Sigma$ becomes zero in the earliest time,
  $M_{\rm gap}$ and $M_{\rm gap,out}$ are the total disc mass and the disc mass outside $r_{\rm gap}$ at $t = t_{\rm gap}$,   
  and $\dot{M}_{*, {\rm gap}}$ is  $\dot{M}_{*}$ at $t = t_{\rm gap}$.
  
\end{table*}

\section{Results}

Results for all simulations are summarized in Table~4. 

\subsection{Discs without dead zones}

\begin{figure}
\begin{center}
\includegraphics[width=84mm]{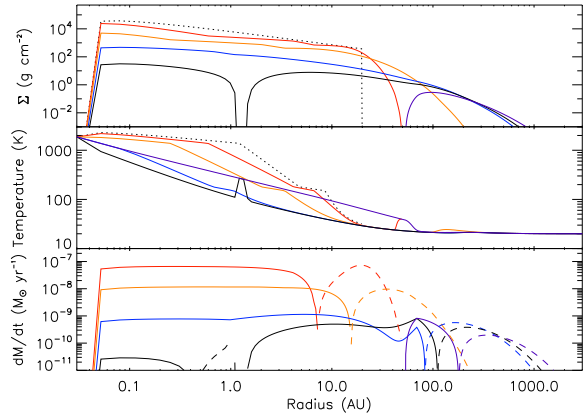}
\end{center}
\caption{
Evolution of surface density, temperature, and mass accretion rate for run N2. 
The viscous $\alpha$ is $10^{-3}$ and is radially constant. 
Black dotted lines are initial conditions. Red, orange, blue, black, and purple solid lines are  values at 
0.1, 1.0, 5.0, 6.01, and 7.0 Myr from the beginning, respectively. 
Mass accretion onto the central star stops when the inner disc disappears at 6.12 Myr.
In the panel of $\dot{M}$, dashed lines represent for outward motion of gas while solid lines for inward motion.} 
\end{figure}

Figure~1 shows time evolution of the surface density $\Sigma$, the temperature $T$, and the 
mass accretion rate $\dot{M}$ for a disc with a radially constant $\alpha$ (run N2).  
As the disc spreads with time, $\Sigma$ and $T$ decrease. 
The kinks seen in the $T$ profile correspond to the opacity transition temperatures (150 K, 180 K, and 1380 K)
and the corresponding kinks are also seen in $\Sigma$.
The direction of the radial gas motion is inward in the inner disc 
and outward in the outer disc,  and the zero radial velocity radius expands with disc expansion. 
The accretion rate $\dot{M}$ is independent of $r$ in the inner disc and is equivalent to  the mass accretion rate onto the central star $\dot{M}_{*}$,
as long as $\dot{M}$ is sufficiently larger than  the mass loss rate due to photoevaporation $\dot{M}_{\rm w}$.
When  $\dot{M}_{*}$ becomes less than $\dot{M}_{\rm w}$, $\Sigma$ rapidly decreases 
and eventually a gap opens at slightly beyond 1 AU, as well as simulations in \citet{Owe11}. 
Once a gap opens, the inner disc (inside the gap) disperses very quickly and mass accretion 
onto the central star stops.  The hole size is only 1.3 AU at this time.
The temperature increases when the disc becomes optically thin, but this phase (with a non-zero gas density)
is very short.  

The gap opening slightly outside $r_{\rm X,in}$ is explained as follows.
Let us assume that the inward mass flux due to viscous accretion at $r_{\rm X,out}$ is  $\dot{M}_{\rm out}$.
As gas moves inward by $dr$, the mass flux is reduced by $2\pi r \dot{\Sigma}_{\rm w}dr$.
Therefore, the inward mass flux decreases with decreasing $r$ and eventually becomes zero at
$r_{\rm X,in}$ if $\dot{M}_{\rm out} = \dot{M}_{\rm w}$.
Thus, if $\dot{M}$ is independent of $r$ without photoevaporation, a gap inevitably opens 
near $r_{\rm X,in}$. Since it takes a time for gas to radially move from $r_{\rm X,out}$ to $r_{\rm X,in}$ 
(the viscous time scale at $r_{\rm X,out}$), $\dot{M}_{\rm out}$ becomes much less than $\dot{M}_{\rm w}$ 
at the time of actual gap opening (see the black curve in Fig.~1).

In some runs, large gaps can open (runs N3 and N6; see Table~4).
This happens because the disc sizes are not sufficiently larger than  $r_{\rm X,out}$ 
and the mass accretion rates are not radially constant near the outer edges of the discs. 
 Even in these cases,  the mass accretion rates onto the central stars 
 at the time of gap opening are much smaller than those seen in transition discs. 
Overall, a large accretion rate and a large hole size are not simultaneously 
reproduced as long as $\alpha$ is radially constant.

\subsection{Discs with dead zones}
\subsubsection{Cases with no dead zone viscosity ($\alpha_{\rm d} = 0$)}

Pictures of evolution of discs with dead zones are very different from those without dead zones. 
Figure~2 shows evolution of a layered disc with $\alpha_{\rm d} = 0$ (run D1).
The mass accretion rate discontinuously drops at $T = 150$ K, because the opacity law changes.
As a result, two dead zone regions appear. Splitting into multiple dead zones was theoretically predicted in 
\citet{Gam96}.
Most of the disc mass remains in the dead zones and $\dot{M}_{*}$ remains nearly constant as long as the dead zones exist. 
Since $\dot{M}$ in the dead zone region increases with $r$, $\Sigma$ increases with time \citep{Gam96, Zhu10}.

\begin{figure}
\begin{center}
\includegraphics[width=84mm]{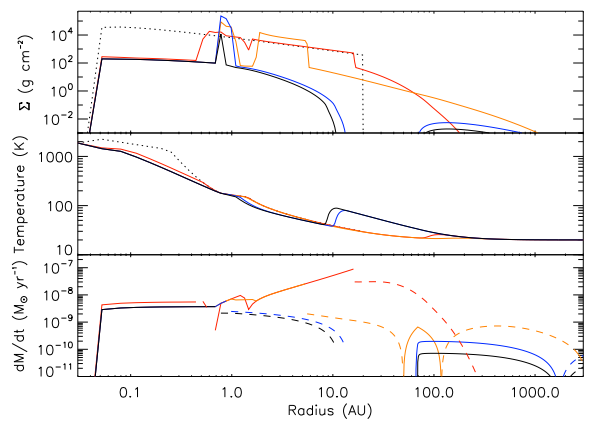}
\end{center}
\caption{Same with Fig.~1 but for run D1. 
A layered accretion is taken into account with $\alpha_{\rm d} = 0$.
Black dotted lines are initial conditions. Red, orange, blue, black solid lines are  values at 
0.1, 2.0, 8.0, and 12.0 Myr from the beginning, respectively. The inner disc completely disappears at 12.17 Myr.} 
\end{figure}

The zero radial velocity radius initially locates at
the outer edge of the outer dead zone.
The direction of radial motion of gas is outward outside the outer dead zone,
while it is inward  in the dead zone region. The absolute magnitudes of the 
mass fluxes in both directions are similar to each other.
Since the size of the dead zone is $\sim 10$ AU, most of gas removal by
photoevaporation is taken place outside the dead zone region.
A discussion of gap opening due to photoevaporation 
is similar to the case of Fig.~1, but now the mass flux is outward. \footnote{Without photoevaporation, 
the direction of radial motion of gas outside the dead zone is outward only in the early expansion phase, but 
eventually turns to be inward \citep{Zhu10}, except near the outer edge of the disc. 
The outward flux is retained by strong photoevaporation in our simulations.}
Thus, a gap opens slightly inside $r_{\rm X,out}$ once 
$\dot{M}$ near the outer edge of the dead zone becomes less than $\dot{M}_{\rm w}$.

For run D1, a gap opens at 48 AU roughly when the outer dead zone disappears, because 
$\dot{M}$ of the outer dead is larger than $\dot{M}_{\rm w}$ 
 while $\dot{M}$ in the inner dead zone is comparable to $\dot{M}_{\rm w}$.
Since the inner dead zone survives for a long time,  $\dot{M}_{*}$ remains high even after a gap opens.
The outward mass flux from the dead zone remains to be  $\sim \dot{M}_{\rm w}$ after a gap opens, as long as the dead zone exists.
Once the dead zone disappears,  the inner disc quickly dissipates and mass accretion onto the central star is halted.

\subsubsection{Cases with finite dead zone viscosity ($\alpha_{\rm d} > 0$)}

\begin{figure}
\begin{center}
\includegraphics[width=84mm]{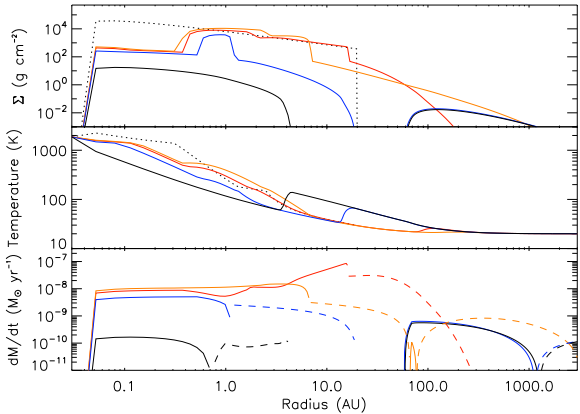}
\end{center}
\caption{Same with Fig.~1 but for run D6. 
A layered accretion is taken into account with $\alpha_{\rm d} = 10^{-5}$.
Black dotted lines are initial conditions. Red, orange, blue, black solid lines are  values at 
0.1, 2.0, 6.0, and 6.22 Myr from the beginning, respectively. The inner disc completely disappears at 6.24 Myr.} 
\end{figure}

\begin{figure}
\begin{center}
\includegraphics[width=84mm]{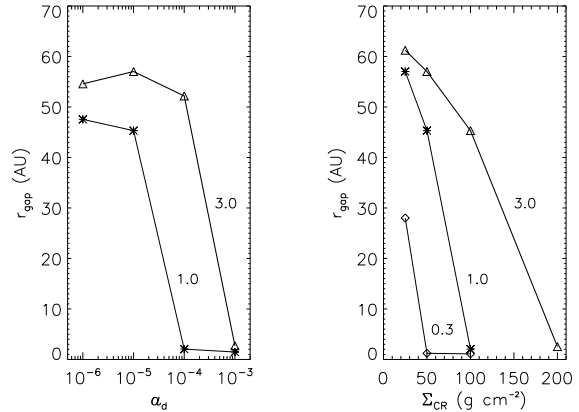}
\end{center}
\caption{Radial locations of gaps. Left: $r_{\rm gap}$ vs.  $\alpha_{\rm d}$ for  $\Sigma_{\rm CR}$ = 50 g cm$^{-1}$
(runs D1, D3, D6, D9, and D15-D18).
Asterisks and triangles are for cases of $L_{\rm X}$ = 1.0 and 3.0 in units of $10^{30}$ erg s$^{-1}$.  
The values of $r_{\rm gap}$ for $\alpha_{\rm d} =0$ are shown at $\alpha_{\rm d} = 10^{-6}$.
Left: $r_{\rm gap}$ vs. $\Sigma_{\rm CR}$ for  $\alpha_{\rm d} = 10^{-5}$ (runs D5-D14). 
 Diamonds, asterisks, and triangles are for cases of 
$L_{\rm X}$ = 0.3, 1.0, and 3.0 in units of $10^{30}$ erg s$^{-1}$, respectively.} 
\end{figure}

It is possible to retain a steady state mass accretion with a finite residual viscosity in the dead zone.
The condition to retain a radially constant $\dot{M}$ is given from Eqs~(\ref{eq:mt}),  
(\ref{eq:nu}),  and (\ref{eq:ald}) as
\begin{equation}
r^{1/2}\frac{\partial}{\partial r} \left[[\Sigma_{\rm a} \alpha _{\rm a} +  (\Sigma -\Sigma_{\rm a})\alpha _{\rm d}] T r^{2}\right] = {\rm const.}
\end{equation}
If the radial profile of $T$ is given such as irradiated discs, the above condition is fulfilled 
when $\Sigma\alpha_{\rm d} \ga \Sigma_{\rm a}\alpha_{\rm a}$ so that the radial profile of  $\Sigma$ is adjusted. 
On the other hand, if viscous heating is the dominant heat source,  
the midplane temperature is given from Eqs.~(\ref{eq:nu}), (\ref{eq:ald}),  (\ref{eq:tveff}), (\ref{eq:tvis}) as 
\begin{equation}
T = \left[\frac{27}{128} \frac{R\kappa_0}{\sigma_{\rm SB}\mu}\Omega
\left((\Sigma^2-\Sigma_{\rm a}^{2})\alpha_{\rm d} + \Sigma_{\rm a}^{2}\alpha_{\rm a} \right)\right]^{1/(3-\beta)}.\label{eq:tdz}
\end{equation}
Thus, if  $\Sigma^2\alpha_{\rm d} \ga  \Sigma_{\rm a}^{2}\alpha_{\rm a} $, the radial profiles of $T$ and $\Sigma$
 are mutually adjusted so that $\dot{M}$ can be independent of $r$.
 
Figure~3 shows an example of evolution of a disc with  $\alpha_{\rm d} = 10^{-5}$ (run D6).
Since this case satisfies the condition $\Sigma^2\alpha_{\rm d} \ga  \Sigma_{\rm a}^{2}\alpha_{\rm a} $,  
$\dot{M}$ becomes almost independent of $r$ (see the orange line at 2.0 Myr in Fig.~3).  Split into multiple dead zones is also suppressed. 
As well as the cases with $\alpha_{\rm d} = 0$, a gap opens at a radius outside the dead zone.

Figure~4 shows the radial location of a gap $r_{\rm gap}$ for various values of  $\alpha_{\rm d}$, $\Sigma_{\rm CR}$, and $L_{\rm X}$.
We define $r_{\rm gap}$ as a radius where 
$\Sigma$ becomes zero in the earliest time while $\Sigma > 0$ in outer radii.
If $\dot{M}$ is  independent of $r$ in the dead zone region, the condition for gap opening is given by $\dot{M}_{\rm w} \ga \dot{M}_{*}$. 
For small values of $\alpha_{\rm d}$ and $\Sigma_{\rm CR}$, $ \dot{M}_{*}$ can be smaller than $\dot{M}_{\rm w}$
even when a dead zone exists. Thus, a gap opens beyond a dead zone in such a case.
On the other hand, for a case with large $\alpha_{\rm d}$ and $\Sigma_{\rm CR}$,
a gap opens at a small radius only after a dead zone disappears and $\dot{M}_{*}$
becomes sufficiently small, as well as the cases with radially constant $\alpha$'s shown in Section~3.1. 
Not surprisingly, gap opening beyond dead zones is possible even with large 
values of  $\alpha_{\rm d}$ and $\Sigma_{\rm CR}$ if $\dot{M}_{\rm w}$ (or $L_{\rm X}$) is large.

\section{Discussion}

\begin{figure}
\begin{center}
\includegraphics[width=84mm]{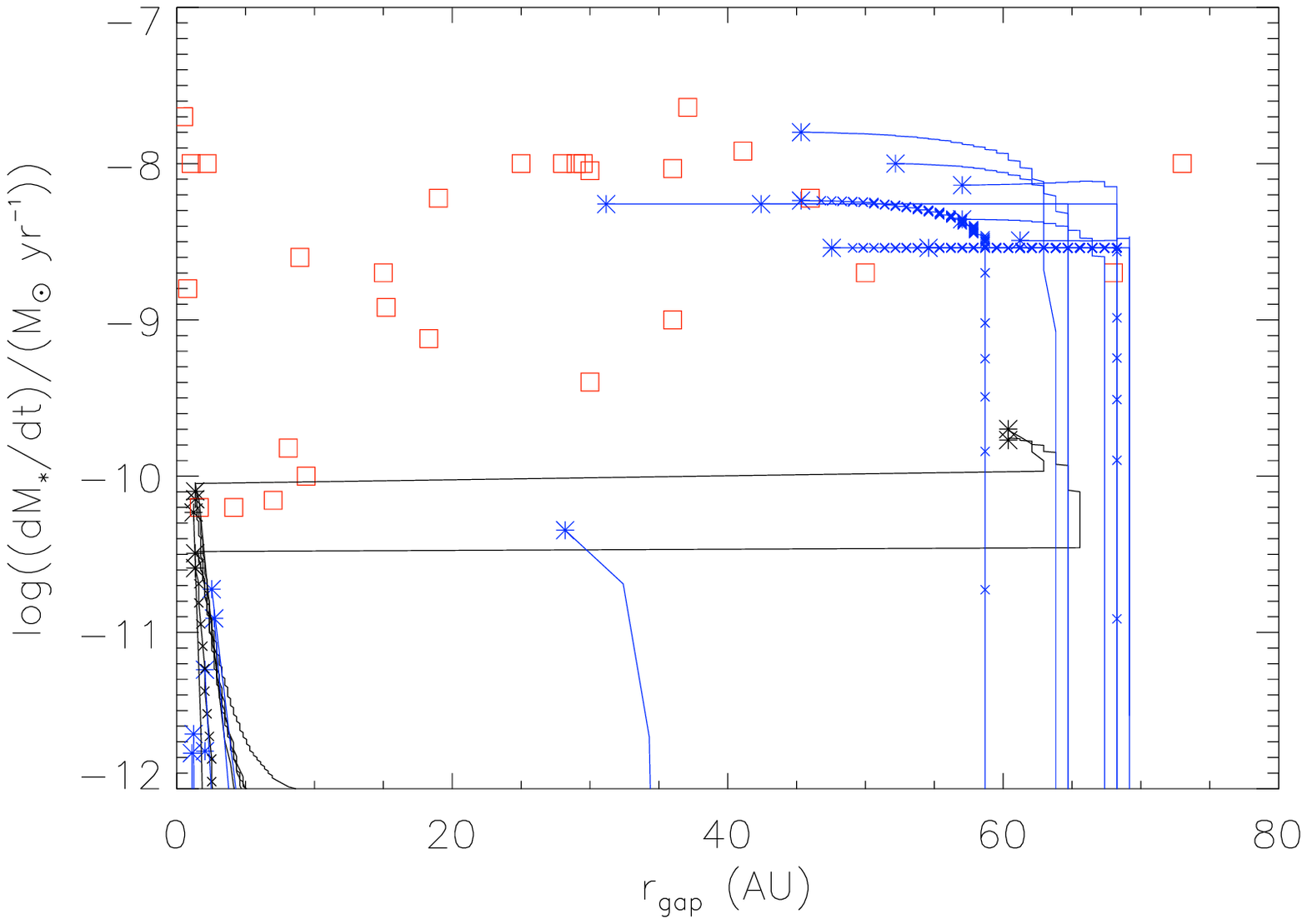}
\includegraphics[width=84mm]{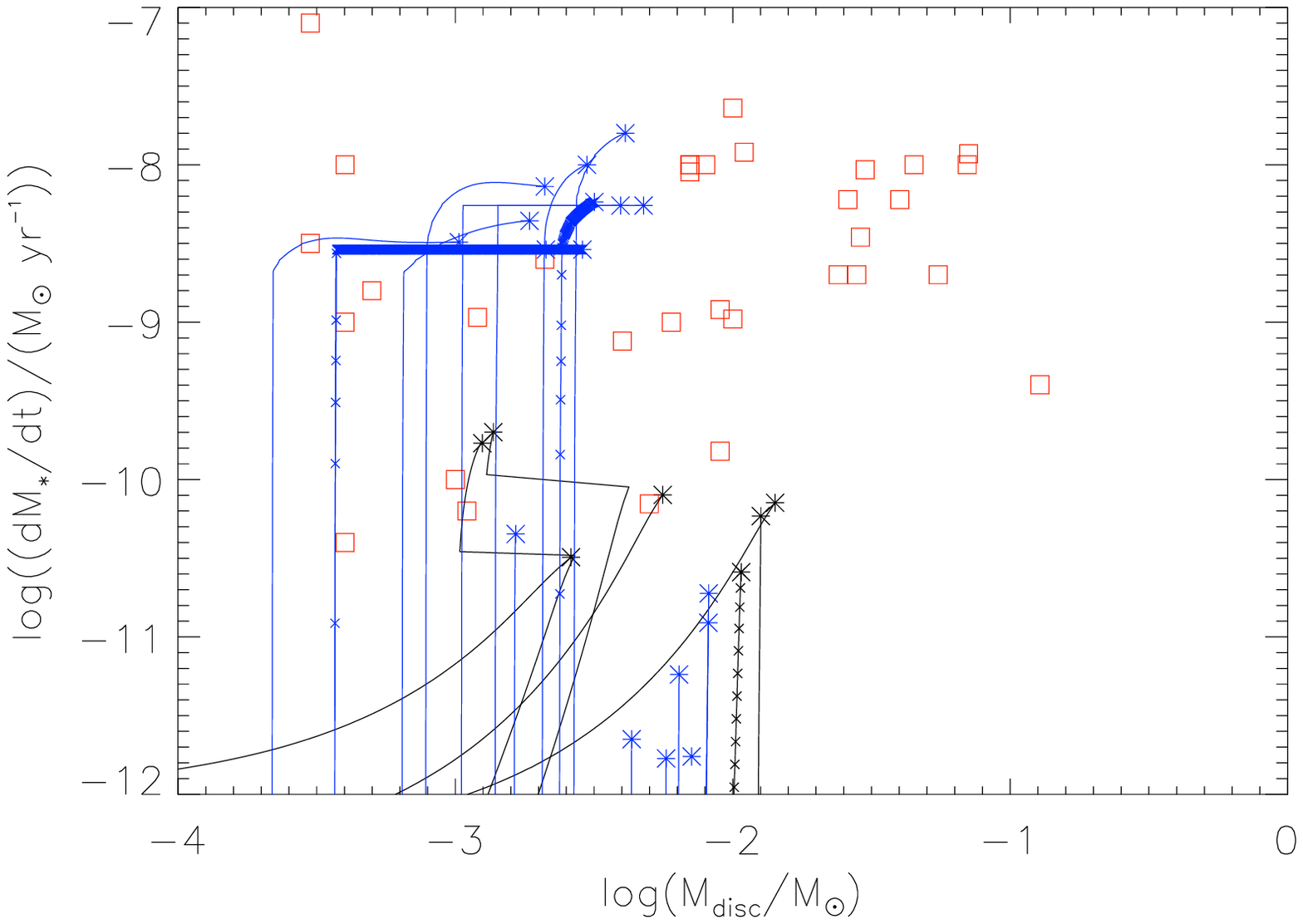}
\end{center}
\caption{Comparison between modeled discs and observed transition discs. Top:  Gap (or hole) size vs. mass accretion rate onto the central star.
Red squares represent observed transition discs,  black and blue asterisks are all simulation results for discs without dead zones
and with dead zones. Subsequent evolution curves are also shown. 
Small crosses represent subsequent evolution of $r_{\rm gap}$ and $\dot{M}_{*}$ with an interval of 10$^{4}$ yr only for runs N2, D1, and D6.
Observed data are from  \citet{Esp08,Esp10}, \citet{Kim09}, \citet{Mer10}, and \citet{And11}. 
Bottom: disc mass vs. mass accretion rate onto the central star. Symbols are the same with those in the top panel. 
For simulations, we take the disc mass $M_{\rm gap,out}$ outside $r_{\rm gap}$, not the total disc mass, $M_{\rm gap}$.
Observational data from \citet{Naj07} are also added.}  
\end{figure}

\begin{figure}
\begin{center}
\includegraphics[width=84mm]{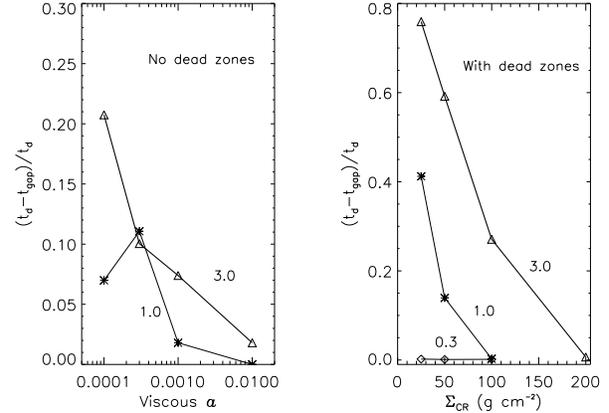}
\end{center}
\caption{Expected fraction of gapped discs for the cases without (left) and with (right) dead zones.
Only discs with mass accretion onto their central stars are considered.  
The time-scales $t_{\rm d}$ and $t_{\rm gap}$ are the times when accretion onto the central star stops and when
a gap opens. The luminosity $L_{\rm X}$  is 0.3 (diamonds), 1.0 (asterisks), and 3.0 (triangles) in units of $10^{30}$ erg s$^{-1}$.
For runs with dead zone, $\alpha_{\rm d} = 10^{-5}$ (runs D5-D14).} 
\end{figure}

Figure~5  shows comparison between observed transition discs and discs from our simulations.
In the upper panel of Fig.~5,  gap sizes and mass accretion rates are plotted.
For simulations, we use $r_{\rm gap}$ and the mass accretion rate onto the central star $\dot{M}_{*, {\rm gap}}$ at the time of gap opening (see Table~4).
Subsequent evolution curves are also shown. 
As can be seen, large gap (or hole) sizes and large accretion rates seen in observed transition discs are well reproduced 
in models with dead zones. On the other hand, models without dead zones are able to 
reproduce only discs with small mass accretion rates and small gap sizes. 
In runs N3 and N6, a gap opens at $\sim$ 60 AU, but the mass accretion rate is small (in these runs, another gap opens
at $\sim$ 1AU, and the disk outside the outer gap quickly dissipates).

In the lower panel of Fig.~5, we compare between disc masses from observations and our simulations. 
For our simulations, we take the disc mass outside the gap  ($M_{\rm gap,out}$ in Table~4), not the total disc mass ($M_{\rm gap}$), 
for consistency with observations.
It should be noted that submillimeter - millimeter observations measure emission from dust primarily around $\sim$ 100 AU
and the surface density of gas is estimated using the interstellar gas-to-dust ratio. 
We find that our model cannot reproduce the masses of the most massive transition discs with high accretion rates,
although the total disc masses including the inner discs in our simulations are comparable to the observed values (see Table~4). 
This means that most of the mass is in the dead zone for a disc with layered accretion.
In our model of layered accretion discs, $\Sigma$ at 100 AU  is  $\sim$ 1 g cm$^{-2}$  before a gap opens but 
it decreases to less than $\sim$ 0.1 g cm$^{-2}$ after a gap opens  (see Figs.~2 and 3). 
On the other hand, observed transition discs show that $\Sigma \sim$ 1-10 g cm$^{-2}$  at 100 AU \citep{And11}.
This value is  similar to or even larger than those seen in classical T Tauri discs \citep{Kit02,And10}. 
The apparent contradiction between modeled and observed disc masses 
might be resolved if dust is accumulated near the inner edge of the outer disc after a gap opens, 
resulting in a large dust-to-gas ratio \citep{Ale07}.
The dust-to-gas ratio may also increase if photoevaporative winds remove only gas but not dust.
 
The expected fraction of gapped discs is 
much larger for discs with dead zones than that for discs without dead zones (Fig.~6), if the disc life times are similar.
The fraction of transition discs in protoplanetary discs 
is more than 50 per cent at 5-8 Myr \citep{Cur11}, and a large fraction of them are accreting \citep{Mer10}.
The fraction of transition discs with large holes ($> 15$ AU) is at least 20 per cent 
among millimeter-bright disc population \citep{And11} although discs with large holes seem to be rarer than those with small holes  \citep{Mer10}.
For discs with dead zones, it is possible to reproduce a high fraction of discs with large holes, 
if  $\Sigma_{\rm CR}$ and $\alpha_{\rm d}$ are small and  $L_{\rm X} $ is large.
A detailed comparison may be able to constrain $\Sigma_{\rm CR}$ and $\alpha_{\rm d}$ with a given distribution 
of $L_{\rm X} $.

Dead zones in our model are optically thick even after gaps open in discs.  
This is inconsistent with observations, although some of transition discs have
 inner optically thick regions indicated from near-infrared  excess (pre-transition discs; \citet{Esp10}).
This indicates that dust removal due to their growth or migration is necessary whereas our current model adopts 
a fixed opacity. Since dust is usually supplied from the outer discs (Brauer, Dullemond \& Henning 2008), gap opening may make the inner discs or dead zones optically thin.
Alternatively, giant planets may clean up dust in the inner discs \citep{Mer10,Zhu11}. 
Giant planets may also be responsible for transition discs with small holes and large accretion rates, as our model does not reproduce such discs (Fig.~5).

\section{Summary}
In this study, we developed a gas disc model which takes into account layered accretion and photoevaporative winds 
induced by X-rays from the central stars. 
We found that a gap opens at a radius outside a poorly-ionized dead zone,
if the mass loss rate due to photoevaporation exceeds the mass accretion rate
in the dead zone region.
Since the dead zone survives even after  the gap opens, 
high mass accretion onto the central star remains for a long time. 

We found good agreements between modeled and observed transition discs in regards to  
gap sizes and mass accretion rates. However, our model shows disc masses (we take masses outside gaps) 
an order of magnitude smaller than those for the most massive observed transition discs. 
This may indicate that the dust-to-gas ratios are large in the outer discs of observed transition discs while our model
assumes a fixed dust opacity.

\section*{Acknowledgments}

We are grateful to anonymous reviewers for fruitful comments on our manuscript.

\label{lastpage}

\end{document}